# Practical Mid-infrared Dual-comb Spectroscopy with Absolute Frequency Referencing Based on a Single-cavity Dual-comb Fiber Laser

Akifumi Asahara[1,*] and Kaoru Minoshima[1,*]

[1] *Department of Engineering Science, Graduate School of Informatics and Engineering, The University of Electro-Communications (UEC), 1-5-1 Chofugaoka, Chofu, Tokyo 182-8585, Japan*

*E-mail: aki.asahara@uec.ac.jp; k.minoshima@uec.ac.jp

We demonstrate a practical mid-infrared (MIR) dual-comb spectroscopy (DCS) method based on a bidirectional dual-comb fiber laser. By exploiting the inherently high relative stability and mutual coherence of the dual-comb laser, we stabilize the MIR dual-comb modes with sufficient precision using only piezoelectric control of the repetition frequencies. This simple configuration enables correction of residual timing jitter and phase noise in interferograms while preserving the absolute optical frequency scale defined by the experimentally obtained comb parameters, without requiring an external continuous-wave laser reference or tight phase-locking. Importantly, the proposed method does not merely produce apparently clean interferograms through signal processing, but enables coherent averaging while retaining both the physically meaningful interferometric waveforms and well-defined absolute frequency information. The proposed system is experimentally validated through MIR spectroscopy of $N_2O$ around 3.9 μm, achieving a signal-to-noise ratio of approximately 690 through coherent averaging of 10,000 interferograms over 65 s, along with the acquisition of an absorption spectrum consistent with the reference data. These results demonstrate a simple and practical approach to absolute-frequency-referenced MIR-DCS.

1. Introduction

Dual-comb spectroscopy (DCS) has attracted considerable attention as a powerful spectroscopic technique in the field of optical frequency combs [1-5]. DCS is an advanced Fourier transform spectroscopy using two optical frequency combs with a slightly different repetition frequency, which achieves unprecedentedly high acquisition speed with excellent precision over a broadband spectral range. DCS has primarily matured in the near-infrared (NIR) region, as the foundational technology for optical combs has been widely established, based on Er fiber laser technology in the NIR. In recent years, advances in coherent wavelength conversion technology have led to its widespread comb application also in the mid-infrared (MIR) region [6-11]. The mid-infrared spectral region contains strong and characteristic molecular absorption features, which are often referred to as molecular fingerprints, and is therefore highly important for a wide range of applications including environmental gas sensing, breath analysis, and plasma diagnostics. Consequently, the development of MIR spectroscopy techniques with high sensitivity, accuracy, and practicality remains an important challenge.

However, conventional DCS systems typically require two independent mode-locked lasers and sophisticated tight phase-locking schemes to synchronize them with high precision. Such complex configurations increase system cost, size, and operational difficulty, which significantly limit the practicality of DCS. To address these issues, dual-comb fiber lasers have been developed mainly in NIR region [12-21], which generate two optical frequency combs from a single laser cavity. Several types of dual-comb fiber lasers have been realized, including two-wavelength [12], bidirectional [13-15], polarization-multiplexed [16,17], and mechanical-sharing types [18-21]. Dual-comb lasers are promising because they enable efficient acquisition of comb spectroscopic information with simple configurations. However, absolute frequency referencing generally remains a challenge in DCS using dual-comb fiber lasers, where some signal processing approach has been applied [20,22,23]. One approach is to employ signal processing with external continuous-wave lasers as frequency reference, which increases the complexity of the system. Alternatively, self-referenced signal correction techniques have also been employed for free-running dual-comb lasers, enabling the acquisition of interferograms with high signal-to-noise ratio (SNR). Such techniques can provide clear interferograms and spectra. However, such signal correction alone does not necessarily guarantee a well-defined and accurate absolute optical frequency scale. Furthermore, such post-processing is generally not performed in real-time operation, making it unsuitable for on-demand measurements or long-term averaging, significantly limiting its practical utility.

In this work, we demonstrate a practical MIR-DCS method based on a bidirectional dual-comb Er-doped fiber laser. The bidirectional dual-comb fiber lasers simultaneously generate two combs propagating clockwise (CW) and counterclockwise (CCW) within a common single-cavity, providing inherently high relative stability and high mutual coherence while simplifying the system configuration. By exploiting these characteristics, we stabilize the MIR dual-comb modes with sufficient precision using only simple piezoelectric control of the repetition frequencies, without requiring an external continuous-wave laser reference or complex tight phase-locking. Furthermore, by developing a real-time interferogram correction scheme, we realize a highly sensitive MIR-DCS with long-term coherent averaging while preserving both physically meaningful interferograms and well-defined absolute frequency information based on experimentally obtained comb parameters. The applicability of this concept is experimentally demonstrated by measuring an absorption spectrum of $N_2O$ gas around 3.9 μm, which is one of the major greenhouse gases. The developed MIR-DCS method offers a simple and practical approach applicable to a wide variety of MIR spectroscopic applications.

## 2. Principle

First, we detail the proposed concept for absolute frequency determination in the MIR-DCS method. As is well known, an optical frequency comb has two frequency parameters: the repetition rate ($f_{rep}$) and the carrier-envelope offset frequency ($f_{ceo}$). The $n$-th longitudinal mode frequency of the comb is determined by $f_n = f_{ceo} + nf_{rep}$. The frequency axis of the comb-resolved spectrum obtained in the DCS is determined based on their frequency differences, $\Delta f_{rep}$ and $\Delta f_{ceo}$. The DCS spectrum is directly acquired in the MHz-GHz radio-frequency region, and is converted into the THz-PHz optical frequency region using a magnification factor of $f_{rep}/\Delta f_{rep}$. Therefore, to obtain correct spectroscopic information with absolute frequencies in the DCS, it is important that all frequency parameters ($f_{rep}$, $f_{ceo}$, $\Delta f_{rep}$, and $\Delta f_{ceo}$) of the two comb sources must be well defined and fixed. This condition is naturally met in a tightly phase-locked DCS. However, in the case of a free-running dual-comb laser, both $f_{\mathrm{rep}}$ and $f_{\mathrm{ceo}}$ of the two combs fluctuate over time. Even if the apparent interferogram and spectrum can be reconstructed through post-processing techniques, the absolute frequency accuracy cannot be guaranteed.

In our proposed method, we employ the following strategy to address this issue in a dual-comb Er fiber laser-based MIR-DCS. First, the two MIR combs are generated using intra-spectrum difference frequency generation (DFG) from the corresponding NIR combs. One key advantage is that the $f_{\mathrm{ceo}}$ of the original NIR comb is canceled, and the generated MIR comb essentially satisfies $f_{\mathrm{ceo}} = 0$. Hence, the optical frequency of the MIR comb mode is determined solely by the $f_{\mathrm{rep}}$. Next, the $f_{\mathrm{rep}}$ must be determined, which is achieved by a simple RF-based stabilization control using a piezoelectric element (PZT) inserted into the fiber cavity. Compared to tight phase-locking methods using electro-optic modulators or ultra-narrow continuous wave laser systems, this approach is much simpler and easier to implement. Although the stabilization accuracy is moderate, the average value of $f_{\mathrm{rep}}$ can be determined. Once $f_{\mathrm{rep}}$ (and $f_{\mathrm{ceo}} = 0$) of the two MIR combs are determined in this way, all frequency components of the MIR-DCS system are uniquely defined, allowing the spectral frequency axis to be correctly determined.

The above method enables us to obtain absolute frequency information using a dual-comb laser. However, even with this stabilization, there remains some residual jitter and phase fluctuations. Therefore, we further develop a real-time signal correction scheme to perform coherent integration as follows (Fig. 1). (i) For each acquired interferogram of DCS, the temporal position of adjacent center burst signals is determined by Gaussian fitting of the envelope. From this information, the fluctuation in the period of the center burst signal ($= 1/\Delta f_{\mathrm{rep}}$) is evaluated. (ii) Next, the time axis is shifted to align the center burst positions. Furthermore, the interferogram data is resampled to match the defined $\Delta f_{\mathrm{rep}}$, which is the reference value in the stabilization control. These processes correct the time axis shift and stretch. (iii) Moreover, the slight residual phase shift within the envelope is corrected to align the phase of each interferogram. (iv) The corrected interferograms are coherently averaged in the time domain. These processes achieve high-SNR coherently integrated interferogram acquisition. Notably, although the method relies on signal processing, all the frequency parameters are determined by the comb light source itself, so it reliably provides accurate absolute frequency information in the optical frequency domain. In addition, by optimizing the analysis program and memory management in the data acquisition device, real-time and long-term integration measurements are realized, rather than post-processing. These features collectively lead to the realization of truly practical MIR-DCS measurements.

## 3. Experiment

The experimental setup for the developed MIR-DCS system is shown in Fig. 2. A lab-built bidirectional dual-comb Er-doped fiber laser was used as the comb light source [15,24]. This laser simultaneously generated two optical frequency combs propagating along CW and CCW directions from a single optical fiber cavity. The mode-locking mechanism was primarily based on nonlinear polarization rotation, and a semiconductor saturable absorber mirror (SESAM) was also used to ensure stable mode-locked operation. The pump laser diode was shared between the two combs, improving their relative balance. The specifications of these comb sources under standard conditions were as follows: the center wavelength and width were 1.55 μm ± 39 nm (CW) and 1.55 μm ± 33 nm (CCW). The output power was 3.8 mW (CW) and 3.0 mW (CCW). The $f_{rep}$ was approximately 40 MHz for both combs. By creating a partially non-common optical path within the cavity and adjusting the position of reflection mirrors, the $f_{rep}$ could be adjusted over a range of several MHz. PZT elements were bonded to the optical fiber in this non-common area, and the $f_{rep}$ of each comb was independently feedback-stabilized by slow servo control referenced to a GPS-disciplined 10 MHz signal. Regarding frequency stability under free-running conditions, the Allan deviations at an averaging time of 1 s were 0.19 Hz for $f_{rep}$ and 0.041 Hz for $\Delta f_{rep}$ (when $\Delta f_{rep}$ = 260 Hz), confirming the high relative stability inherent to the dual-comb laser. With the PZT-based feedback control, the Allan deviation of $\Delta f_{rep}$ was further reduced to 0.0005 Hz.

MIR comb generation was achieved using the following single-pass configuration [24]. First, the dual-comb laser output was amplified by an Er-doped fiber amplifier (EDFA), and the spectral bandwidth was broadened to cover approximately 1.0–2.0 μm using a highly nonlinear optical fiber (HNLF). The broadened comb was then focused into a waveguide-type periodically poled lithium niobate crystal (WG-PPLN) [25] designed to meet the phase matching conditions of DFG. As a result, a broadband MIR comb was generated for both CW and CCW combs based on the principle of intra-comb DFG. The generated MIR comb covered a wavelength range of 3.2–4.4 μm, and the output power of this MIR component exceeded 200 μW. The spectral shape and power were highly dependent on the state of broadened light incident on the WG-PPLN. Thus, by adjusting the parameters such as excitation power for EDFA and polarization state before the EDFA, the generated MIR comb could be optimized for measurement targets.

MIR-DCS was performed as follows. The two MIR combs emitted from the WG-PPLN were spatially overlapped using a beam splitter after removing the near-infrared (NIR) component with a Ge optical long-pass filter. The combined MIR dual-comb was then coupled into an $InF_3$ single-mode fiber, which matched the spatial modes and improved the visibility of interference. The MIR dual-comb was passed through an optical band-pass filter in the MIR band as required for the measurement, and it was then transmitted through a $N_2O$ gas cell (10 cm long, 100 torr) as the sample. The interference signal was detected using a high-speed MCT detector with 100 MHz bandwidth. The output RF signal was passed through a 20-MHz RF low-pass filter to match the Nyquist range of the DCS system, and it was recorded using a high-speed digitizer with 100 MHz bandwidth and 14-bit resolution. A lab-built LabVIEW-based measurement application was installed on the digitizer control system, enabling simultaneous data acquisition and real-time signal analysis.

## 4. Results and Discussion

Figure 3 shows an interferogram acquired with the MIR-DCS system. The $\Delta f_{rep}$ was set to 155 Hz so that the DCS measurement spectral range sufficiently covered the $N_2O$ absorption band around 3.9 μm. Under single-shot conditions, an interferometric waveform with an SNR of approximately 11 was obtained. Although coherence degradation is a potential concern due to phase noise introduced during the wavelength conversion processes, the present system provided interferograms that were sufficiently observable. However, in this case, weak free induction decay (FID) responses of $N_2O$ molecule were obscured by noise. After applying the proposed real-time signal correction and coherent averaging, the interferogram quality was dramatically improved. As also shown in Fig. 3, coherent averaging over approximately 10,000 interferograms, corresponding to an acquisition time of about 65 s, increased the SNR to approximately 690. We confirmed that the SNR was improved in proportion to the square root of averaging time. Notably, the peak intensity of the signal was maintained nearly constant even after the averaging process. These behaviors clearly indicate that timing jitter and phase fluctuations were effectively compensated by the correction scheme, enabling high-quality long-term coherent averaging. At this high SNR, the FID signal originating from $N_2O$ molecules became clearly observable.

MIR spectra were obtained by Fourier transformation of the coherently averaged interferograms. First, spectra acquired without the gas cell exhibited good agreement with those obtained using a commercial optical spectrum analyzer, confirming correct spectral acquisition by the MIR-DCS. When the $N_2O$ gas cell was introduced, characteristic absorption lines were clearly observed, as shown in Fig. 4. The measured line positions showed good agreement with the HITRAN database, demonstrating that the proposed dual-comb laser-based MIR-DCS system provides reliable and well-defined absolute frequency information without complex and tight phase-locking system.

By comparing the center frequencies of the absorption lines obtained by the MIR-DCS with those listed in the HITRAN database, the absolute frequency accuracy was estimated to be approximately 2 GHz (0.1 nm). This accuracy is sufficient for identifying molecular absorption lines in various practical applications, although further improvement is required for precision spectroscopy. One of the main limitations arises from uncertainty in the estimation of $\Delta f_{rep}$ fluctuations using interferogram envelope fitting ((i) in Fig. 1). This scheme requires sufficiently stable single-shot interferograms with adequate SNR. In the present measurement, the mutually coherent bidirectional dual-comb laser system enabled the acquisition of single-shot interferograms with an SNR of approximately 11, allowing the proposed real-time correction method to function effectively. However, the limited single-shot SNR still restricts the correction accuracy, and further improvement in absolute frequency accuracy is expected by enhancing the interferogram quality. The SNR of the MIR interference signal can be improved through optimization of the light source, such as increasing the amplified optical power and improving the robustness of the system against environmental disturbances. In addition, near-infrared (NIR) interferograms with high SNR can be readily acquired from a separate output port of the dual-comb laser. Because these NIR interferograms are temporally correlated with the MIR interferograms, hybrid coherent processing using the NIR signal as a timing reference is expected to further improve the frequency accuracy of MIR-DCS.

## 5. Conclusion

We demonstrated a practical MIR-DCS method based on a bidirectional dual-comb fiber laser. By exploiting the inherently high relative stability and mutual coherence of the dual-comb laser, absolute frequency referencing was achieved without an external continuous-wave laser or tight phase-locking, by combining the generated MIR combs with simple piezoelectric stabilization of the repetition frequencies. Furthermore, real-time signal correction and long-term coherent averaging enabled the acquisition of high-SNR MIR interferograms while preserving both physically meaningful interferometric waveforms and well-defined absolute frequency information based on experimentally obtained comb parameters. The proposed system was experimentally validated through MIR-DCS of $N_2O$ around 3.9 μm, achieving an SNR of approximately 690 through coherent averaging of 10,000 interferograms over 65 s, along with an absorption spectrum consistent with the HITRAN database. The proposed approach offers a simple and practical solution for absolute-frequency-referenced MIR-DCS and is expected to be applicable to a wide range of MIR spectroscopic applications.

**Acknowledgments**

This research was supported by MEXT/JSPS KAKENHI Grant Numbers 21H05014 and 26H02248. It was also supported by JST ERATO MINOSHIMA Intelligent Optical Synthesizer Project (JPMJER1304). We sincerely thank Dr. Y. Nakajima, Dr. K. Yoshii, Dr. H. Tian, Dr. T. Kato, Mr. J. Li, Mr. G. Fukawa, Mr. T. Shimizu, and Mr. U. Yamada for their contributions to the technical foundation of this work.

# Figures

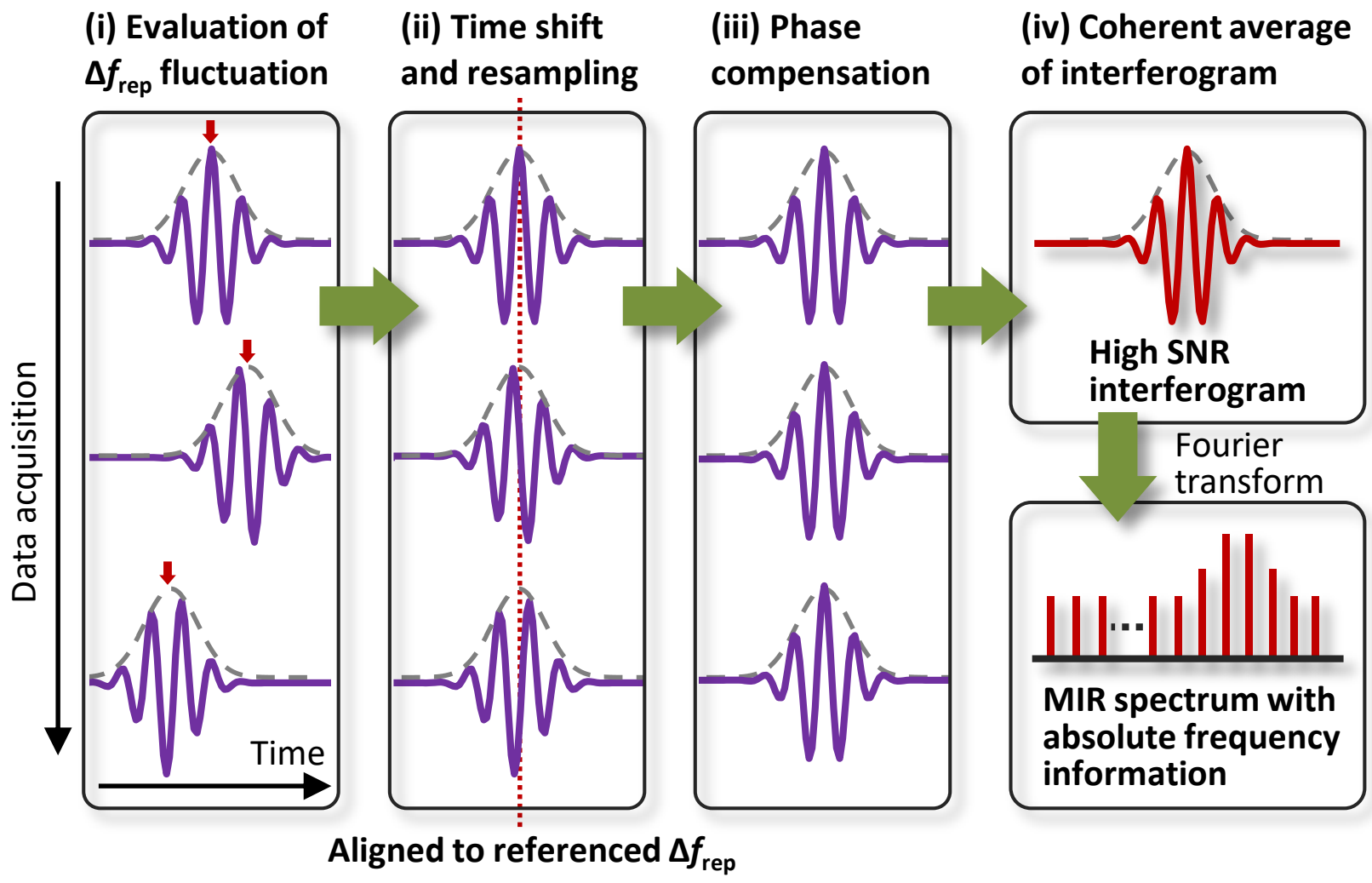

**Fig. 1.** Proposed concept of coherently averaged MIR-DCS with absolute frequency information using a dual-comb laser without complex phase locking.

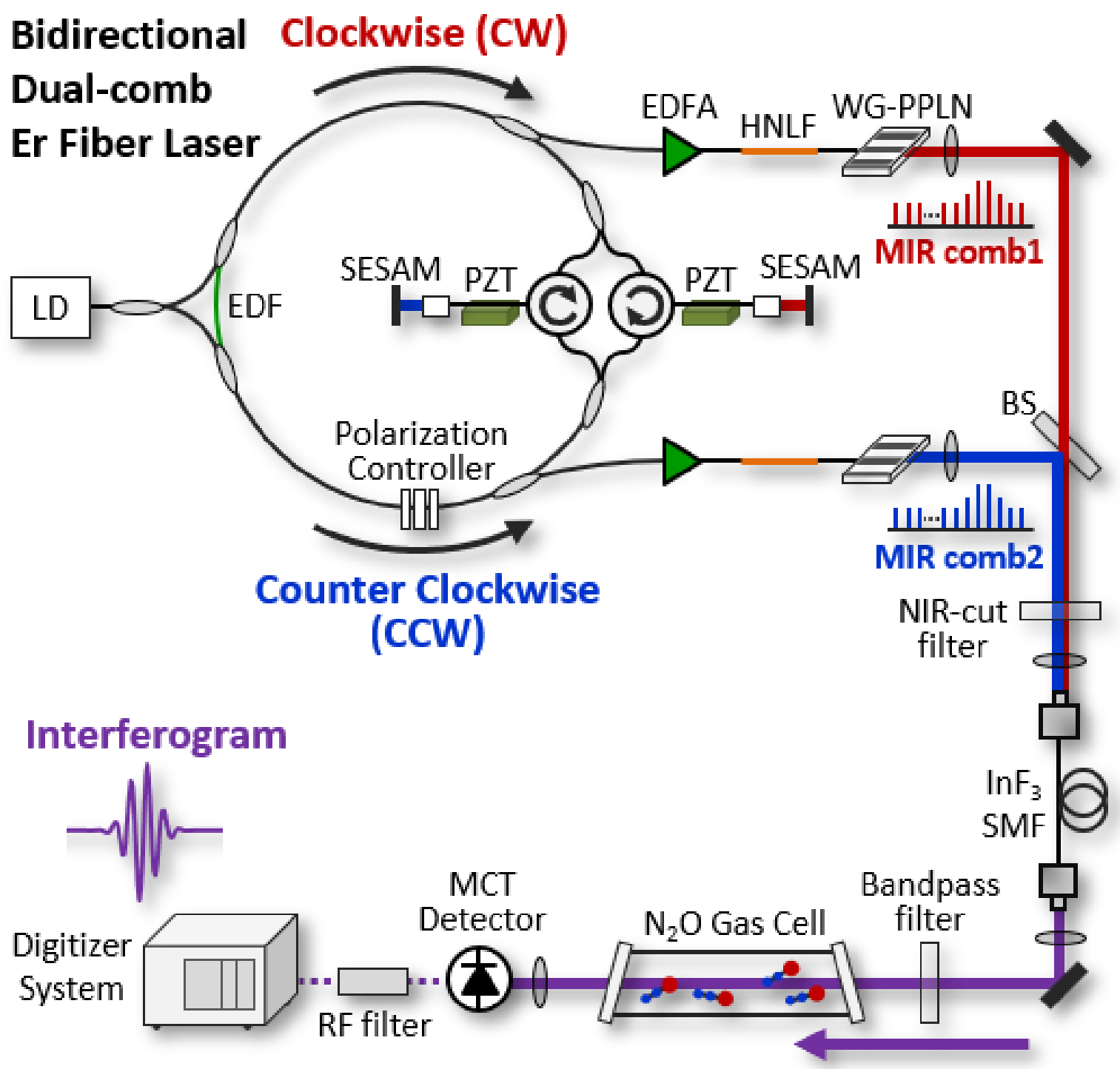

**Fig. 2.** Experimental setup of MIR-DCS using a bidirectional dual-comb Er fiber laser. SESAM: semiconductor saturable absorber, EDFA: Er-doped fiber amplifier, HNLF: highly nonlinear fiber, WG-PPLN: waveguide-typed periodically-poled lithium niobate.

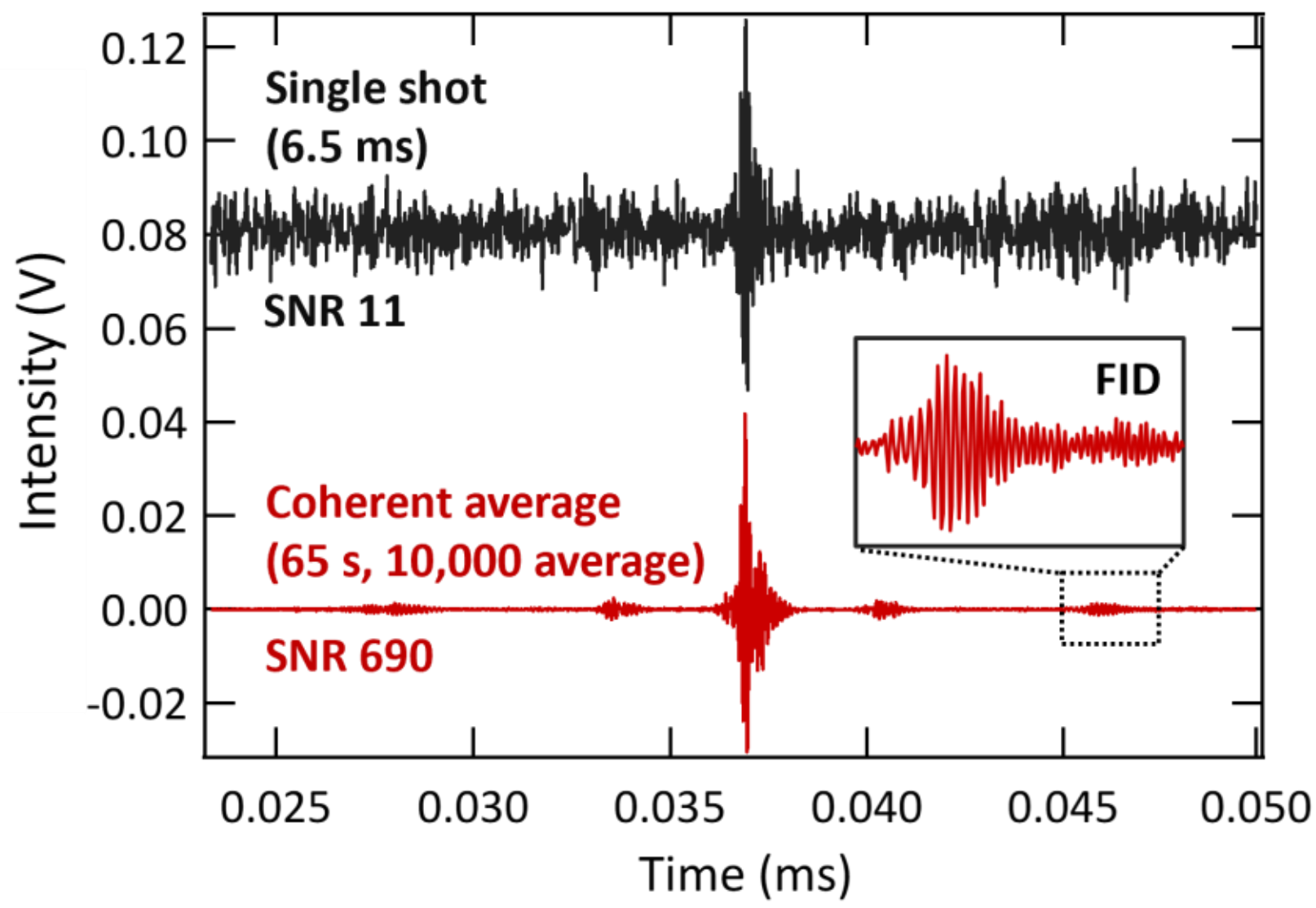

**Fig. 3.** MIR-DCS interferograms acquired under single-shot (black) and coherently averaged (red) conditions. The single-shot signal is vertically offset for clarity. The FID signal observed in the coherently averaged interferogram is shown in the inset on an enlarged scale.

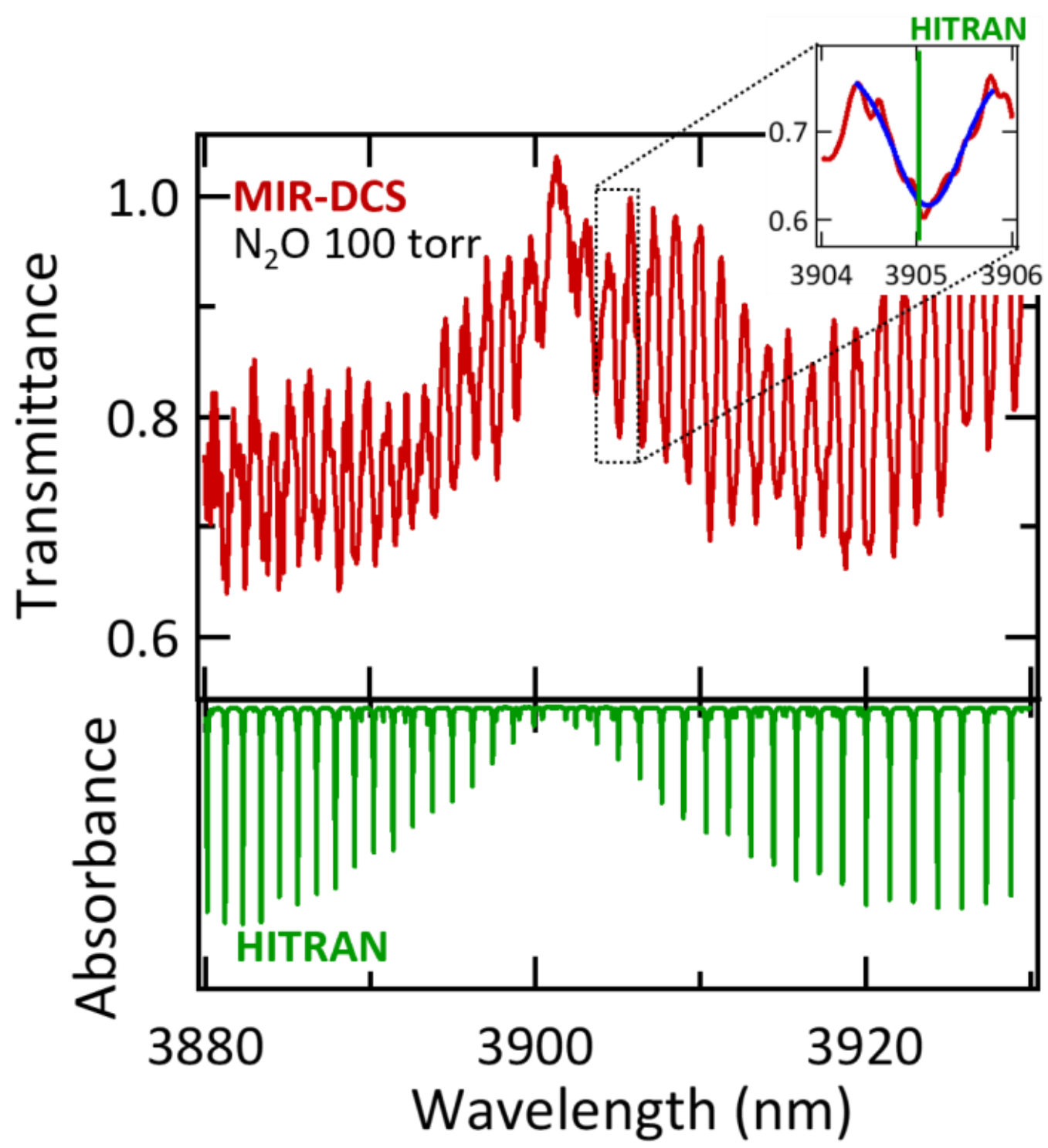

**Fig. 4.** MIR-DCS spectrum of $N_2O$ gas (red) and corresponding absorption lines from the HITRAN database (green). The inset shows an enlarged view of a representative absorption line, comparing the measured spectrum with the HITRAN reference.